\documentclass[showpacs,aps,prb,twocolumn,groupedaddress]{revtex4-1}

\usepackage{graphicx}
\usepackage{multirow}

\def\Brperp{Br$_2^\perp$\ }
\def\Brpar{Br$_2^\Vert$\ }

\begin{document}


\title{Bromination of Graphene and Graphite}


\author{A. Yaya$^1$}
\author{C. P. Ewels$^{1}$}
\email{Corresponding Author : chris.ewels@cnrs-imn.fr}
\author{I. Suarez-Martinez$^{1,2}$}
\author{Ph. Wagner$^1$}
\author{S. Lefrant$^1$}
\author{A. Okotrub$^3$}
\author{L. Bulusheva$^3$}
\author{P. R. Briddon$^4$}
\affiliation{$^1$ Institut des Mat\'eriaux Jean Rouxel, Universit\'e de Nantes, CNRS UMR 6502, 44322 Nantes, France}
\affiliation{$^2$ Nanochemistry Research Institute, Curtin University of Technology, Perth, WA 6845, Australia}
\affiliation{$^3$ Nikolaev Institute of Inorganic Chemistry SB RAS, 3 Acad. Lavrentiev avenue, Russia}
\affiliation{$^4$ School of Natural Sciences, University of Newcastle upon Tyne, Newcastle upon Tyne, UK}

\date{\today}

\begin{abstract}
\textbf{For full reference please see: Phys. Rev. B 83, 045411, 2011}\\
\\
We present a density functional theory study of low density bromination of graphene and graphite, finding significantly different behaviour in these two materials.  On graphene we find a new Br$_2$ form where the molecule sits perpendicular to the graphene sheet with an extremely strong molecular dipole.  The resultant Br$^+$-Br$^-$ has an empty p$_z$-orbital located in the graphene electronic $\pi$-cloud.  Bromination opens a small (86meV) band gap and strongly dopes the graphene.  In contrast, in graphite we find Br$_2$ is most stable parallel to the carbon layers with a slightly weaker associated charge transfer and no molecular dipole.  We identify a minimum stable Br$_2$ concentration in graphite, finding low density bromination to be endothermic.  Graphene may be a useful substrate for stabilising normally unstable transient molecular states. 
\end{abstract}


\maketitle


The intercalation of carbon nanomaterials with electron donors and acceptors is an active research area in which much effort is channelled towards the understanding and controlling of the electronic properties of graphene and graphite. Numerous potential applications such as sensors, electronic display panels, hydrogen storage and supercapacitors \cite{Seung-Hoon2002} have been suggested for such intercalated materials.  The layered structure of graphites plays an important role in charge transfer reactions.  Acceptor species can intercalate between graphitic layers, expanding the graphite with the resultant hybrids known as graphite intercalated compounds \cite{Dresselhaus2002}.
   
  Bromine acts as an acceptor when intercalated in materials such as graphite or nanotubes, and has been proposed experimentally as a way to open a band gap in 3- or 4- layer graphene\cite{Jung2009}. The in-
plane electrical conductivity of graphite increases from $2.4 \times 10^{4}\ \Omega ^{-1} $cm$^{-1}$ at room temperature to  $2.2 \times 10^{5}\ \Omega ^{-1} $cm$^{-1}$ after intercalation with bromine\cite{Dresselhaus2002}.

Bromine forms many ordered phases in graphites and undergoes an order-disorder phase transition as the
amount of bromine or temperature changes \cite{Dresselhaus2002,Bandow2002,Yoichi2003,Eeles1964, Bach1963,Bardhan1980}. Bromine intercalated graphite forms
stage-$n$ compounds where $n$ is the number of graphitic layers between planes of Br$_2$ ($n>1$ \cite{Sasa1971}). Extended X-ray Absorption Fine Structure (EXAFS) spectroscopy at
room temperature and 100K \cite{Heald1778} showed that intercalated bromine molecules lie parallel to the basal
plane, with an expansion of the Br-Br distance by 0.03 \AA\ to accommodate the lattice mismatch
between the free molecule and the 2.46 \AA\ spacing between graphite hexagons. X-ray diffraction and
electron microscopy studies \cite{Eeles1964} suggest that intercalated Br$_2$ at lower concentration is composed of
chains of Br$_2$ molecules in which the intermolecular distances is identical to that of solid bromine.

Graphite Raman spectra associated with Br$_2$ intercalation show a strong peak at 242 cm$^{-1}$ \cite{Ecklund1978, Erbil1983,Erbil1982,Defang2007} assigned to the intercalated
Br$_2$ stretch mode. The frequency is downshifted from 320 cm$^{-1}$ for gaseous Br$_2$ and 295 cm$^{-1}$ for solid
molecular bromine\cite{Cahill1966}.  There have been limited density functional studies of 
brominated graphite\cite{Widenkist2009} and graphene\cite{Seung-Hoon2002,Rudenko2009}.

In this paper we examine low density bromination of graphene and graphite using density
functional (DFT) calculations within the local density approximation\cite{Briddon2000}. The method has
been successfully used to study intercalated boron in graphite\cite{Suarez-Martinez2007}. A localised Gaussian basis set is used with a large
number of fitting functions per atom (22 for each C atom and 50 for each Br), with angular
momenta up to $l$=2 for C and $l$=3 for Br. A finite temperature electron level filling of kT=0.04eV is
used to improve convergence. Core electrons were eliminated using norm-conserving relativistic
pseudopotentials of Hartwigsen, Goedecker and Hutter\cite{Hartwigsen1998}. A cut-off energy of 150 Hartrees was
used to obtain convergence of the charge density. 

Isolated Br$_2$ was calculated in a 13.23 \AA\ cubic supercell. Hexagonal $4\times4$ graphene supercells
containing C$_{32}$Br$_2$ were used with a large vacuum spacing of 31 \AA\ between layers to ensure no inter-
layer interaction, and a $4 \times 4 \times 1$ Monkhorst-Pack k-point grid\cite{Monkhorst1976}. Graphite calculations used $3 \times 3 \times n$
layer supercells (C$_{18}$)$_n$(Br$_2$)$_m$, $n$=1-4, $m$=1-2) for different layer stackings, with $4 \times 4 \times 1$ or $4 \times 4 \times 2$ k-point grids depending on cell size.
All structures were fully geometrically optimised with no constraints of symmetry, allowing both
atomic positions and cell dimensions to vary freely. Atomic charge states were obtained by
summing Mulliken population analysis over all the filled electronic states. Vibrational frequencies
were calculated by determining the energy and forces for $\pm$ 0.2 au displacements of the Bromine
atoms. The second derivatives on the displaced atoms can then be found from the two-sided
difference formula for the second derivative.  All results are spin averaged, test calculations with spin polarisation all gave zero spin solutions as the most thermodynamically stable.


    
Our calculated bond length, 2.29\AA, and stretching frequency, 326 cm$^{-1}$, for isolated Br$_2$ show excellent
agreement with experiment (2.27\AA\ and 323 cm$^{-1}$\cite{Ecklund1978}, 2.283\AA\ and 320 cm$^{-1}$ \cite{Erbil1982}) and literature DFT/LDA values (2.263\AA \cite{Miao2009}, 2.244\AA\ and 324 cm$^{-1}$ \cite{Seung-Hoon2002}).

\begin{figure}
\begin{tabular}{cccc}
(a) 0.00 & (b) +0.02 & (c) +0.05 & (d) +0.11 \\
\includegraphics[width=1.7cm]{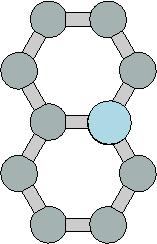} & \includegraphics[width=1.7cm]{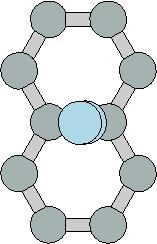} & 
\includegraphics[width=1.7cm]{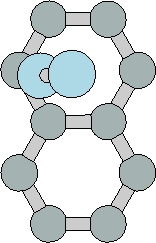} & \includegraphics[width=1.7cm]{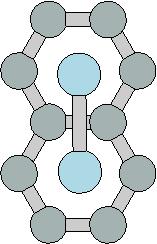}  \\
(e) +0.14 & (f) +0.14 & (g) +0.15 & (h) +0.15 \\
\includegraphics[width=1.7cm]{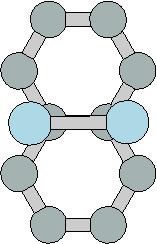} & \includegraphics[width=1.7cm]{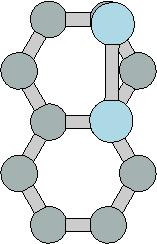} & 
\includegraphics[width=1.7cm]{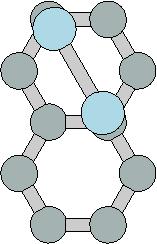} & \includegraphics[width=1.7cm]{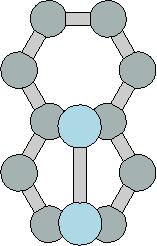}
\end{tabular}
\caption{Optimised geometries of Br$_2$ molecule perpendicular (a-c) and parallel (d-h) to a
graphene sheet. Relative stabilities are quoted in eV (binding energy of Br$_2$ in structure (a) is 0.40eV). The text 
refers to (a) \Brperp and (d) \Brpar \label{brominepic}}
\end{figure}


 \begin{table}
 \caption{Calculated parameters for the most stable \Brperp and \Brpar on graphene compared with literature DFT calculations.\label{strucparams}}
 \begin{ruledtabular}
 \begin{tabular}{lcc|c}
 & \Brperp  & \Brpar & Literature \\ \hline
 Orientation & Perp. & Parallel & Parallel \\
 Binding energy / Br$_2$ (eV) & -0.40 & -0.29 & -0.20\cite{Seung-Hoon2002}\\&&& -0.29\cite{Rudenko2009} \\
 Br-Br Stretch Frequency (cm$^{-1}$) & 288 & 270&311\cite{Seung-Hoon2002}\\
 Br-Br bond length (\AA) &2.33&2.31&2.245\cite{Seung-Hoon2002}\\
 Br-C distance (\AA) & 2.74&3.45&3.375\cite{Seung-Hoon2002} \\&&&3.74\cite{Rudenko2009} \\
 Charge state of Br atoms (e) 
&+0.48 & -0.04&  \\
&-0.61&-0.04 \\
Total Charge transfer / Br$_2$ (e) &
-0.13 & -0.08& 
 \end{tabular}
 \end{ruledtabular}
 \end{table}

 
Figure \ref{brominepic} shows the structures of Br$_2$ over graphene in both perpendicular and parallel
orientations after geometry optimisation, with calculated properties for \Brperp and \Brpar(structures (a) and (d) respectively in Figure \ref{brominepic}) in Table \ref{strucparams}. 

The most thermodynamically stable arrangement is Br$_2$ oriented perpendicular to the graphene sheet above a
carbon atom (Figure \ref{brominepic}a, referred to hereafter as \Brperp). 
Its binding energy of 0.40eV shows it will be strongly
physisorbed at room temperature. However the small variations in binding energy
between structures suggests that low density Br$_2$ binding to graphene will be largely orientation
independent; indeed and at these densities at room temperature Br$_2$ should be in constant tumbling motion
(this also holds for the results of\cite{Rudenko2009}). 
Increasing Br$_2$ concentration did not significantly change the relative energies of perpendicular and parallel
orientations. However it is possible to obtain twice the maximum
surface density for \Brperp as for Br$_2$ in the parallel orientation (Figure \ref{brominepic}d, referred to hereafter as \Brpar).
Thus for the limit of high surface concentrations we expect \Brperp to dominate.

\Brperp represents a very unusual configuration for bromine.  
It shows strong charge transfer (0.129e) from the graphene, with a very strong induced molecular dipole (Br$^{+0.480}$-Br$^{-0.609}$).  The singly occupied p$_z$-orbital of the lower Br atom depopulates into the p$_z$ of the upper Br atom 
forming a nascent Bromonium and Bromide ion pair.  In this way the emptied lower p$_z$-orbital can sit 
within the graphene $\pi$-cloud (Br only 2.74 \AA\ above the graphene). 

This behaviour is reminiscent of the well-known reaction between Br$_2$ and unsaturated bonds in organic chemistry.  However in these cases this
dipolar form of Br$_2$ is an unstable transient state that immediately saturates the C=C bond, forming two Br-C bonds. In graphene this final step would be endothermic due to steric hindrance between
the Br atoms as a result of the mechanical confinement of the lattice.  Indeed our attempts to stabilise C-Br pairs on graphene in both neighbouring (1,2) and cross-hexagon (1,4) configurations both resulted in Br spontaneously reconstructing into a Br$_2$ molecule (placing Br-C in a (1,4) configuration with Br atoms on opposite sides of the graphene is 1.68eV less stable than \Brperp).

\begin{figure}
(a) \includegraphics[width=5.0cm,angle=-90]{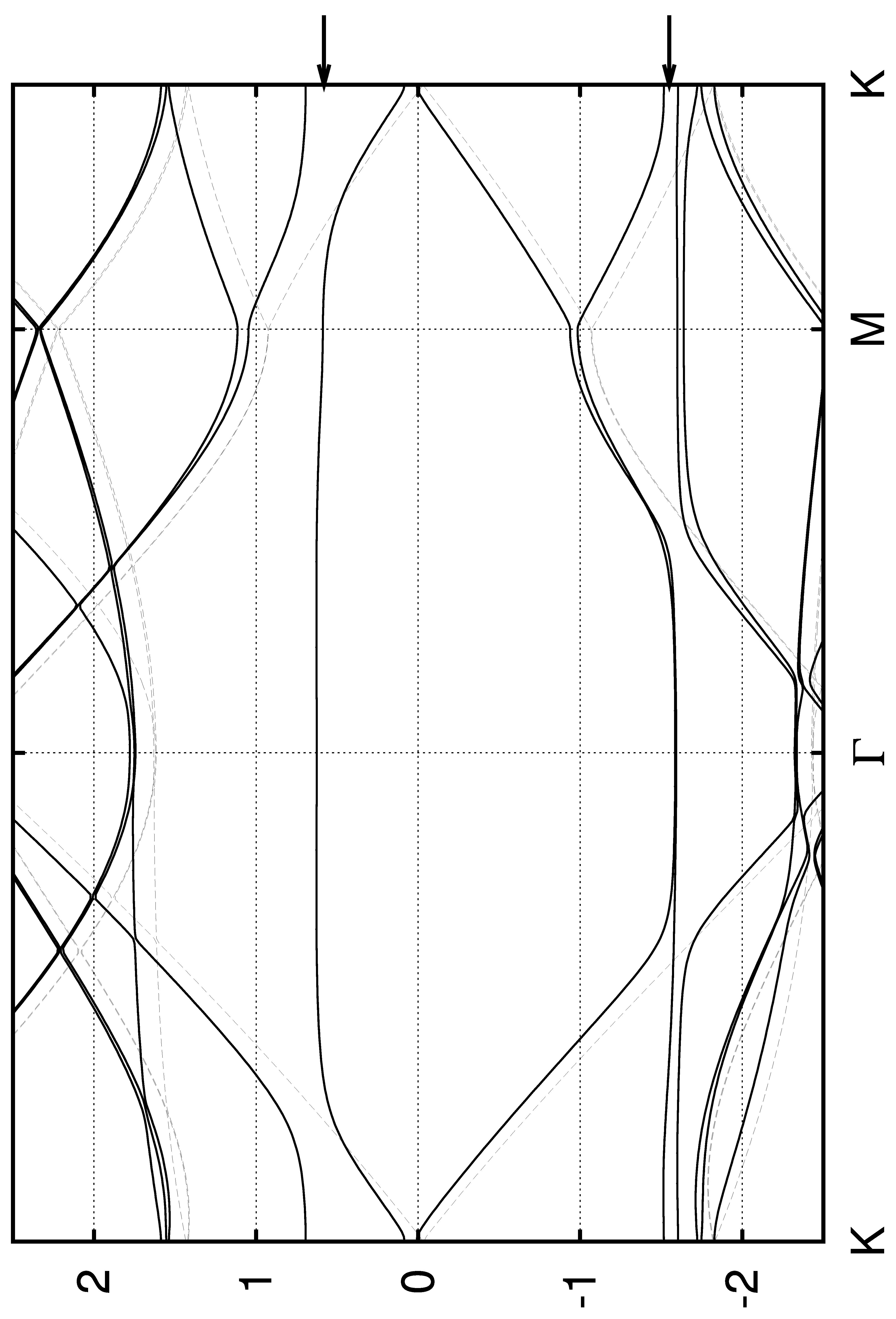} \\
(b) \includegraphics[width=5.0cm,angle=-90]{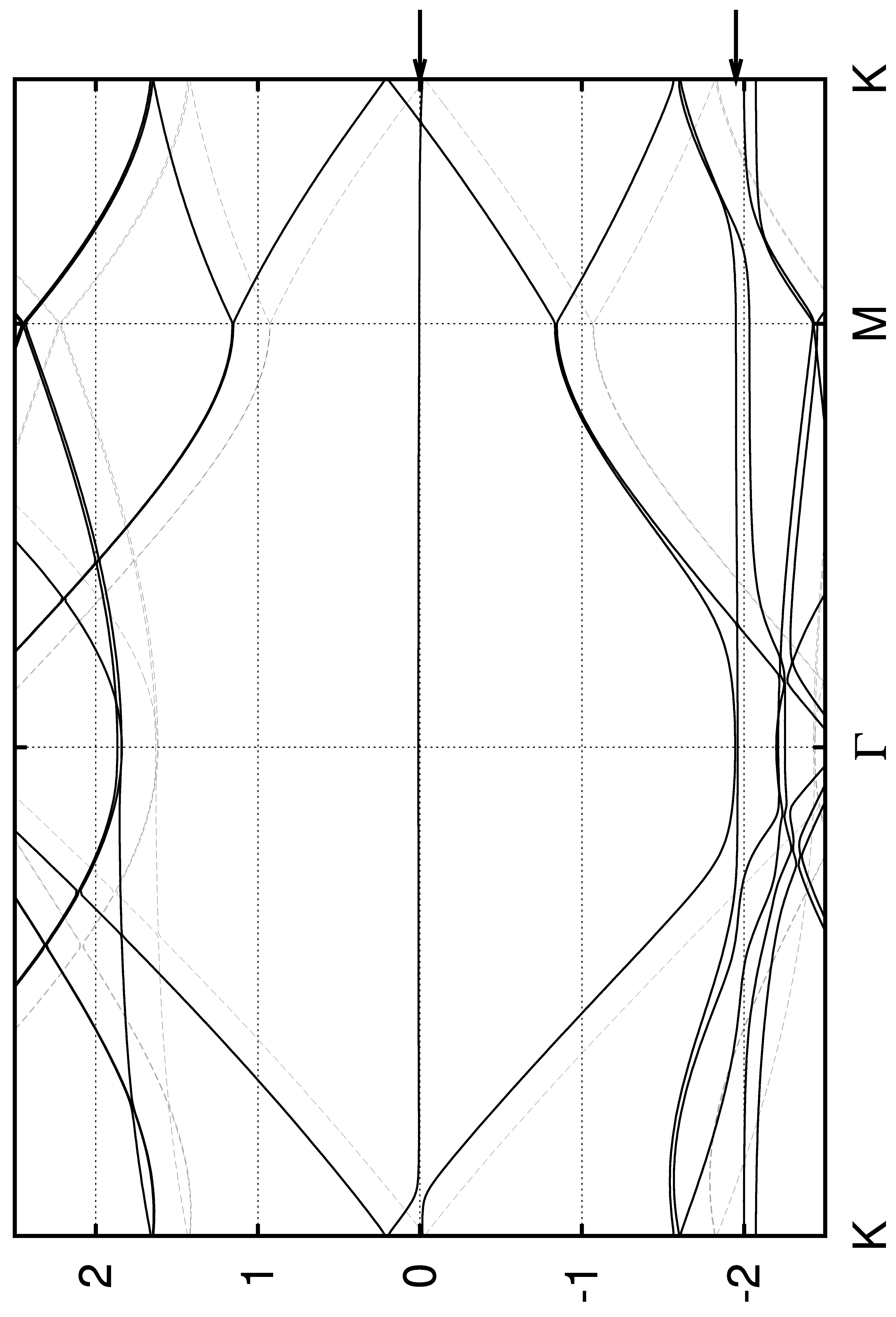} 
\caption{Band structure of (a) \Brperp and (b) \Brpar on graphene (eV), faded dotted lines indicate the same supercell of pristine graphene for comparison.  Arrows indicate bromine related states. \label{bandstructures}}
\end{figure}

The unusual \Brperp configuration is reflected in the band structure (Figure \ref{bandstructures}a). 
The strong coupling between the graphene LUMO and the Br$_2$ antibonded state at around +0.6 eV reflects the interaction between the empty Br p$_z$-orbital and the graphene $\pi$-cloud, with the resultant low density Br$_2$ layer
opening a small 86meV band gap. 

In contrast \Brpar has no induced dipole, with weaker charge transfer (0.084e) from the graphene. The molecule sits
3.254 \AA\ above the graphene {\it i.e.} above the $\pi$-cloud, with the additional charge occupying the 
Br$_2$ $pp\sigma^{*}$ anti-bonding state. The band structure (Figure \ref{bandstructures}b) shows the
Br$_2$ states lie lower than those of \Brperp by $\sim$0.7eV.
The Br$_2$ anti-bonded state pins the Fermi level $\sim$0.2eV lower than in the pristine case, leaving graphene states around the K-point depopulated, indicating charge transfer from graphene to Br$_2$. 
The Bromine states are flat and largely decoupled from the graphene bands since
there is only weak interaction between Br$_2$ orbitals and the graphene $\pi$-cloud in this orientation.


Our results are consistent with experimental Raman observations of brominated graphene\cite{Jung2009}.
While for graphene with three or more layers a resonant Raman signal was observed for Br$_2$ at
around 240cm$^{-1}$, for mono- and bi-layer graphene no bromine signal was observed. This could be
an orientation effect, since if the Br$_2$ sits perpendicular to the surface as we propose and orthogonal
Raman is used, then the molecules will be aligned with the beam and there will be no interaction
and hence no signal. 
In addition the bromine HOMO/LUMO states are further apart for \Brperp ($\sim$2.2eV) than for \Brpar ($\sim$1.96) 
or the various graphite structures we have examined (1.7-2.0eV). Given the excitation laser used
(633nm, 1.96eV), \Brperp may not be in resonance as the authors suggested \cite{Jung2009}. 
\Brpar would have an associated Raman signal, and since none is observed this allows us to exclude this configuration. 
We note that the strong dipole of \Brperp will make the molecule infra-red active.

We now turn our attention to graphite.
Our calculated energy to separate AB graphite layers is 36.7 meV/atom, with AA stacked
graphite 12.0meV/atom less stable than AB stacked graphite. These figures are in good
agreement with experiment (35meV/atom\cite{Benedict1998}) and previous calculations (9.68-9.70meV/atom
AA/AB energy difference\cite{Telling2003,Charlier1994}). Our interlayer spacings of 3.39\AA\ and 3.50\AA\ 
for AB and AA-stacked graphite respectively are also in reasonable agreement with previous calculations\cite{Charlier1994}.

We placed Br$_2$ in a variety of different orientations and locations including above $\alpha$- and $\beta$- carbon
atom sites and hexagon centers, in graphite of various layer stackings. 
Unlike graphene, Br$_2$ in graphite is more stable parallel
to the graphitic layers with Br atoms above hexagon centers (see Table \ref{graphiteparams}), in
agreement with experiment\cite{Eeles1964,Heald1778}.
Perpendicular oriented Br$_2$ structures are much less stable (by typically 0.52eV). 
The results are summarised and compared with experiment in Table\ref{graphiteparams}.  
We note that the charge transfer value from Raman was reported with a large uncertainty\cite{Erbil1982}.

 \begin{table}
 \caption{Calculated and experimental results for Br$_2$ intercalated graphite. 
 The C-C layer distance refers to the
layers separated by bromine.  Experimental values from EXAFS \cite{Heald1778}, Raman \cite{Ecklund1978,Erbil1982} and XRD \cite{Eeles1964,Erbil1983,Sasa1971,Rudorff1941}. 
At these low concentrations Br$_2$ intercalation is endothermic. \label{graphiteparams}}
 \begin{ruledtabular}
 \begin{tabular}{l|ccc}
 & Stage-1 & Stage-2 & Experiment \\
Cell used &  C$_{18}$Br$_2$ & C$_{72}$Br$_4$ & \\
Binding Energy/Br$_2$ (eV) & +0.08 & +0.08 & \\
C-C layer distance (\AA) & 6.45 & 6.47 & 7.0\cite{Sasa1971,Erbil1983} 7.05\cite{Rudorff1941}  \\
$c$-axis (\AA) & 9.80 & 9.82 & 10.7\cite{Eeles1964} \\
Br-Br bond length (\AA) & 2.30 & 2.30 &  2.34 \cite{Heald1778,Heald1980}\cite{Erbil1983} \\
Br-C distance (\AA) & 3.51 & 3.46 & 2.9\cite{Heald1778} \\
Br-Br frequency (cm$^{-1}$) & 287 & 274 & 242-258 \cite{Ecklund1978,Erbil1982} \\
Charge transfer / Br$_2$ (e) & -0.10 & -0.12 & -0.16\cite{Heald1980} -0.34 \cite{Erbil1982} \\
 \end{tabular}
 \end{ruledtabular}
  \end{table}

Besides the lowest energy structures quoted here we obtained many metastable structures.
While their energies were all within 0.01-0.05eV of
those structures discussed here, they show significant variation in the Br$_2$ stretch frequency (250-278cm$^{-1}$), and slight variation in position and Br-Br bond length (2.31-2.33\AA). This suggests that at room temperature Br$_2$ in graphite will be mobile, and is consistent with the observation
of a broad and somewhat complicated Raman peak \cite{Ecklund1978}.

We explored all possible layer stacking combinations for stage-1 and stage-2 intercalated graphites.
Bromine molecules are most stable with AA-stacked graphite each side, while unbrominated graphite layers 
preferentially are AB stacked.
Thus (indicating layers of bromine molecules by X) we find the most stable stage-1 phase to be [AX]$_n$,
and stage-2 to be [AXABXB]$_n$, suggesting [AXAB]$_n$ and [AXABABXBAB]$_n$ for
stage-3 and stage-4 respectively.  

Our calculated bromine intercalation energy is weakly endothermic, since at these low densities the
energetic cost associated with separating graphite planes (a cost per unit area) is not sufficiently
offset by the binding energy of Br$_2$ to the layers. 
Subtracting the energy to separate graphite layers from the Br$_2$ intercalation energy
gives an energy for intercalation of Br$_2$ into `pre-separated stage-1 graphite' of 0.581eV/Br$_2$.
This implies that the {\it minimum} Br$_2$ concentration for exothermic intercalation in stage-$n$ graphite will be C$_{16n}$Br$_2$. Indeed a fixed-Br calculation for a C$_8$Br$_2$ stage-1 high coverage structure gives Br$_2$ intercalation as exothermic.  Thus Br$_2$ will aggregate in the same inter-layer space and should be considered as a layer rather than individual molecules.

This minimum required concentration also suggests intercalation will be a slow diffusion process with an
abrupt diffusion front. This is consistent with long experimental intercalation times \cite{Fischer2002},
despite the high bromine inter-layer mobility. It also explains why, 
on out-gassing bromine, the material switches from a stage-2 to compound to
stage-$n$ ($n$=3, 4, ...) rather than remaining stage-2 with lower bromine density per layer\cite{Dresselhaus2002, Ecklund1978,Fischer2002}.


We note that frequency calculations incorporating the energetic double derivatives of surrounding carbon atoms gave identical values to within 1 cm$^{-1}$, showing that the Br$_2$ stretch mode is decoupled
from the surrounding carbon lattice consistent with the literature\cite{Seung-Hoon2002}. 

Band structure calculations of Br$_2$ layers between graphite sheets (not shown here) give a bromine-related state which pins the Fermi level just below that of perfect graphite ($\sim$0.1eV) indicating charge transfer from graphite to Br$_2$.  This state shows some dispersion indicating weak coupling with the underlying graphite. In other respects the graphite band structures are barely perturbed.

Our results on graphite and graphene can explain the anomalously large G peak shift for single graphene sheets in comparison with few-layer graphene\cite{Jung2009}. 
Since the higher maximum Br$_2$ surface density for \Brperp than for inter-layer
Br$_2$ means charge transfer
per unit area will be higher. Additionally Br$_2$ can attach to both sides of graphene.
We find a binding energy of -0.38eV /Br$_2$ for two
\Brperp either side of the same C atom, with associated charge transfer of -0.12e/Br$_2$.  Thus the net total charge transfer per
unit area will indeed be significantly higher for monolayer graphene than multi-layer systems.

Test calculations for a (5,5) single walled nanotube, either isolated or in bundles, gave similar structural behaviour, {\it i.e.} Br$_2$ on the surface of the isolated tube adopts a
perpendicular orientation, while intercalated Br$_2$ sits parallel to the tube walls. This will
be explored further in a later publication.

In summary, we have examined low density Br$_2$ adsorption in graphene and graphite.
 On graphene Br$_2$ adopts an unusual perpendicular orientation, opening a small band gap ($\sim$86meV) in the
graphene with strong charge transfer. 
The molecule forms a Br$^{+}+$Br$^{-}$ pair, rendering it infra-red active. This is a new form of Br$_2$ previously only considered as an unstable intermediate to bromine induced carbon bond saturation.  Such graphene-induced stabilisation behaviour may be mirrored in other molecular species, enabling study of otherwise unstable reactive molecular forms.

In graphite Br$_2$ adopts a parallel orientation to the sheets with an associated charge transfer. Our
calculations are in good agreement with experimental data where available. At high bromine
concentrations and low temperatures there is some evidence of bromine chain structure formation
\cite{Chung1986,Chung1977} in graphite, and we are currently investigating this further.
We note that high density bromination of graphite leads to stage-2 compounds, and in conjunction
with an appropriate secondary surfactant this may be a promising way to produce bilayer graphene.

\begin{acknowledgments}
CPE and PW thank the ``NANOSIM\_ GRAPHENE'' project
ANR-09-NANO-016-01 funded by the French National Agency (ANR) within the P3N2009 programme.
\end{acknowledgments}


\bibliographystyle{apsrev4-1}

\end{document}